# Disparity in the Evolving COVID-19 Collaboration Network

Huimin Xu[1], Redoan Rahman[1], Ajay Jaiswal[1], Julia Fensel[2], Abhinav Peri[2], Kamesh Peri[3], Griffin M Weber[4], and Ying Ding[1]*

[1] School of Information, University of Texas at Austin, Austin TX 78701, USA
{huiminxu, redoan.rahman, ajayjaiswal, ying.ding}@utexas.edu
[2] Westlake High School, Austin TX 78746, USA
{ Julia.smag1, abhinavperi16}@gmail.com
[3] Katana Graph, Austin TX 78701, USA
kamesh.peri@katanagraph.com
[4] Harvard Medical School, Boston MA 02115, USA
griffin_weber@hms.harvard.edu
*Corresponding author

**Abstract.** COVID-19 pandemic has paused many ongoing research projects and unified researchers' attention to focus on COVID-19 related issues. Our project traces 712,294 scientists' publications related to COVID-19 for two years, from January 2020 to December 2021, in order to detect the dynamic evolution patterns of COVID-19 collaboration network over time. By studying the collaboration network of COVID-19 scientists, we observe how a new scientific community has been built in preparation for a sudden shock. The number of newcomers grows incrementally, and the connectivity of the collaboration network shifts from loose to tight promptly. Even though every scientist has an equal opportunity to start a study, collaboration disparity still exists. Following the scale-free distribution, only a few top authors are highly connected with other authors. These top authors are more likely to attract newcomers and work with each other. As the collaboration network evolves, the increase rate in the probability of attracting newcomers for authors with higher degree increases, whereas the increase rates in the probability of forming new links among authors with higher degree decreases. This highlights the interesting trend that COVID pandemic alters the research collaboration trends that star scientists are starting to collaborate more with newcomers, but less with existing collaborators, which, in certain way, reduces the collaboration disparity.

## 1    Introduction

The science of science is a field to study the structure and evolution of science, which has offered rich quantitative and qualitative methods to uncover hindsight about crea-tivity, collaboration, and impact in scientific endeavors. Despite the prominent contributions of science of science researchers, which are deeply rooted in normal science including scientific collaboration (Leahey, 2016), team composition (Wu et al., 2019), novelty (Uzzi et al., 2013), and funding allocation (Jacob & Lefgren, 2011), studies about scientific activities in abnormal conditions are largely overlooked. However, patterns or findings from studies on normal science cannot be applied to abnormal conditions. Normal science was coined by Thomas Samuel Kuhn (1962) as a phase of science during which the scientific community has confidence in what the world is like. Normal science often suppresses fundamental differences/novelties because it favors fitting phenomena into the widely accepted conceptual theories/boxes (Collins, 1994). So, when a novel pandemic occurs, we need to understand how scientists collaborate and what the team dynamics are, whether out-of-the-box thinking can be supported by scientific communities. This paper studies the scientific collaboration of COVID-19 authors from the perspective of evolving networks.

The ongoing COVID-19 pandemic certainly not only disturbed the normal routines of scientific activities, but also demanded solutions from science to resolve the spread. Understanding scientific activities in abnormal times are urgent and imperative (Fry et al., 2020). Studying the patterns of scholarly communications during pandemic times can help us understand how science can bend the trajectories of pandemic spreading and provide implications for science policy makers to have a better risk management plan for future unexpected disasters.

## 2   Related Work

Barabasi et al. (2002) explored the evolving collaboration networks in mathematics and neuro-science disciplines covering eight years. They found that the average degree (i.e., degree centrality) increases and the node separation (i.e., the average distances of all shortest paths between two given nodes) decreases. In addition to uncovering the power law distribution of networks (Barabási & Albert, 1999), he also revealed two mechanisms to explain the preferential attachment phenomenon – "the richer get richer". In the research, Barabasi et al. (2002) found that a new author is more likely to work with authors who already have many coauthors. Also, authors who already have many authors are more likely to build more links as the network evolves. Azondekon et al. (2018) analyzed the connectedness of researchers in malaria research by building a co-authorship network from papers collected by Web of Science. They found that prolific authors have higher probabilities of collaborating with more authors and the giant component covers 94% of all the vertices to confirm a small-world network. Furthermore, Uddin et al. (2013) extended the relationship between network centrality measures with the impact and productivity of authors. They established the regression model and revealed that degree centrality and betweenness centrality of authors are positively correlated with the strength of their scientific collaborators (i.e., number of coauthors of a given author) and impact (i.e., the citation count of a research article authored by a given author). In our project, we want to apply the network science measures into the COVID-19 collaboration network to detect the collaboration disparity during the pandemic

## 3   Methodology

**Data**

We use the LitCOVID dataset[1] as our source of COVID-19 publications. LitCOVID collects COVID-19 publications from PubMed dataset[2] through searching relevant keywords, such as "coronavirus", "ncov", and "2019-nCoV" (Chen et al., 2020; Chen et al., 2021). The results are updated daily and reviewed by human and machine learning algorithms. By December 23rd in 2021, there are 205,476 COVID-19 papers, including specific PubMed id and title information. By tracing the PubMed id in the PubMed dataset, we can get author lists and publication time of each paper. We deal with the author name disambiguation problem with the assistance of Semantic Scholar dataset[3] (Ammar et al., 2018). Xu et al. (2020) evaluated the effect of author name disambiguation in the Semantic Scholar, which reaches 96.94% in F1 score. LitCOVID and Semantic Scholar both keep the PubMed id, thus we can match author names in LitCOVID with unique author ids in Semantic Scholar through the common PubMed id. Finally, we get 186,046 COVID-19 related papers, with complete publication time, author names, and author ids from 2020 January to 2021 December. Among these papers, there are 712,294 unique authors. A majority of 89% papers (166,126) papers have more than one author and 99% (704,164) authors have collaborators. In order to observe the evolution of the collaboration network, we document eight quarters based on publication time. Table 1 and Figure 1 describe the cumulative nodes and links added into the network over time. We can see that the node increase rate appears to be stable, whereas the link grows suddenly at the first and second quarter in 2021 and then keeps stable.

**Table 1. Cumulative number of nodes and links for the COVID-19 collaboration network up to a given time**

| Date | N of nodes | N of links |
| --- | --- | --- |
| 2020_Q1 | 13,062 | 139,562 |
| 2020_Q2 | 128,589 | 2,223,421 |
| 2020_Q3 | 245,330 | 11,310,419 |
| 2020_Q4 | 349,479 | 16,507,786 |
| 2021_Q1 | 469,920 | 85,371,561 |
| 2021_Q2 | 573,649 | 146,582,029 |

---

[1] https://www.ncbi.nlm.nih.gov/research/coronavirus/
[2] https://pubmed.ncbi.nlm.nih.gov/download/
[3] https://api.semanticscholar.org/corpus/download/

| | | |
|---|---|---|
| 2021_Q3 | 658,173 | 162,823,954 |
| 2021_Q4 | 712,294 | 177,493,364 |

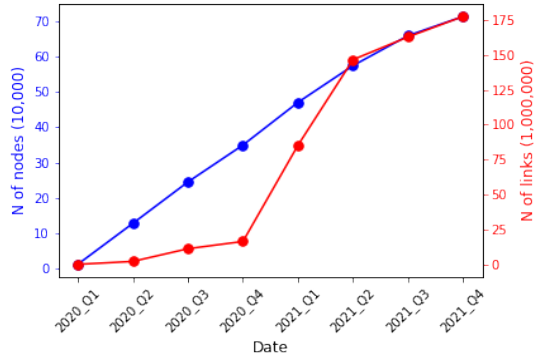

**Figure 1. Cumulative number of nodes (left: indicating authors) and links (right: indicating co-authorship) for the COVID-19 collaboration network up to a given time**

**Measures**
Authors have formed the collaboration network if any given two authors have co-authored at least one paper, there is an edge to connect these two nodes. Degree centrality for an author $i$ can be defined as: $Degree\ Centrality\ (a) = \frac{k}{n-1}$, where $k$ is the degree of author $a$ (represents the number of authors with whom author $a$ is directly connected in the co-authorship network), n represents the number of authors in the network.

Fig 2 shows how to calculate the probability of attracting external new nodes and forming new internal links. Probability of attracting new authors for an old node with degree $k_i$: $P(k_i) = \frac{V(k_i)}{N(k_i)}$, where $V(k_i)$ means the number of newcomers that authors with degree $k_i$ attract, $N(k_i)$ means the number of authors with degree $k_i$. Probability of forming new links among old nodes with degree $k_i$ and $k_j$: $P(k_i, k_j) = \frac{L(k_i,k_j)}{N(k_i)*N(k_j)}$, $L(k_i, k_j)$ means the number of new links between authors with degree $k_i$ and $k_j$, $N(k_i) * N(k_j)$ means the number of combination pairs between authors with degree $k_i$ and $k_j$.

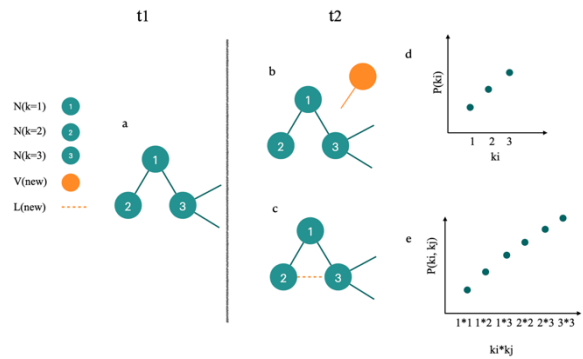

**Figure 2. The illustration of calculating probability of attracting new nodes and forming new links among old nodes. a. At the time t1, there are three kinds of nodes with different degrees (green) in the network. We calculate the number of these nodes, like, N(k=1) means the number of nodes with degree 1. At the next time t2, new nodes (orange) join in (b) and new links (orange) among old nodes appear (c). V(new) represents the number of new nodes, and L(new) represents the number of new links. d. We calculate the probability of attracting new nodes for an old node with degree $k_i$. f. We calculate the probability of forming a new link for an old node with degree $k_i$ and an old node with degree $k_j$.**

## 4   Result

We observe the evolution of the collaboration network at eight different stages, from the first quarter of 2020 to the fourth quarter of 2021. Firstly, we found the degree distributions of networks up to the indicated time all follow scale-free power law distribution (Fig 3). We can see most of the authors have a relatively small number of collaborations, but a few authors have the ability to connect with many partners. This result is consistent in these eight networks. The degree distributions gradually shift upward as more new authors join the community, and meanwhile move rightward as existing authors enhance their collaboration ties.

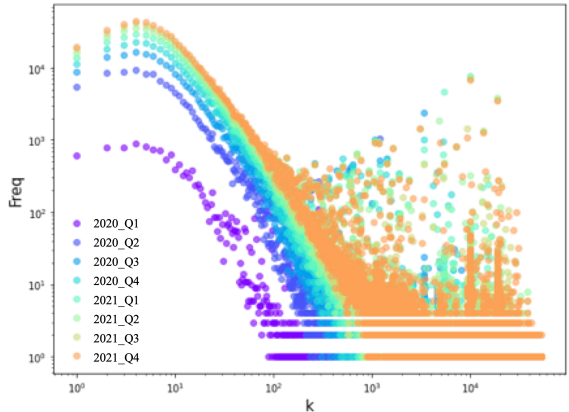

**Figure 3. Degree distribution for COVID-19 authors, showing the cumulative results up to a given time. The plot is drew using a logarithmic scale for both the x-axis and the y-axis. X axis k represents the number of collaborators an author has, whereas y-axis frequency shows the number of authors in the network with degree k.**

Given the scale-free distribution, we choose to separately observe the top and tail authors' degree centrality. The tail 20% authors' average degree centrality values have trivial changes over time (Fig 4a). At the very beginning (2020_Q1), the degree centrality is relatively large for top 10% and top 20% authors as the network has a few nodes. We also need to note that one year later (2021_Q1), there is an apparent increase in degree centrality for top authors. Meanwhile, the gap in degree centrality between top 10%, top 20% and tail 20% increases in 2021. These patterns suggest that top authors play a more important role in connecting other authors than tail authors, which increases the collaboration disparity. We are also curious about what kind of collaborators top authors connect with. In the first year, the difference in degree centrality between top authors' collaborators and tail authors' collaborators is evident. It indicates that they prefer to work with homogenous authors whose degree is similar to them. Specifically, top authors tend to work with top authors, whereas tail authors tend to work with tail authors. When the COVID-19 pandemic suddenly starts, the powerful alliance among top authors enables them to react quickly to the outbreak. However, one year later, the difference in degree centrality between top authors' collaborators and tail authors' collaborators is less significant (Fig 4b). One possible reason is that although top authors still work with each other, they attract more newcomers as more people pay attention to the COVID-19 pandemic and join the community. Thus, high degree centrality and low degree centrality cancel out each other.

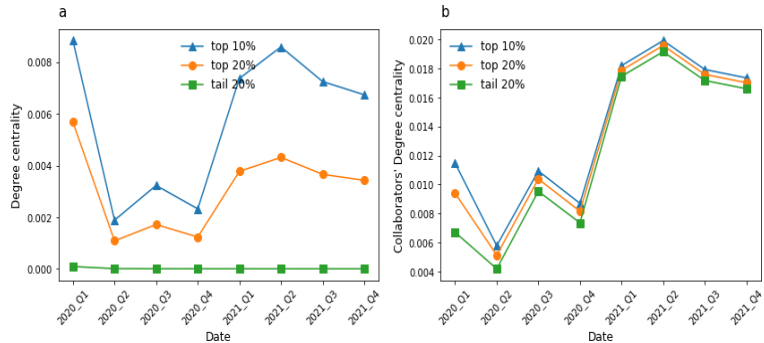

**Figure 4. a. Average degree centrality values of COVID-19 authors whose degree centrality values are in the top 10%, top 20% and tail 20%. b. Average degree centrality values of top 10%, top 20% and tail 20% authors' collaborators.**

To explain the phenomenon above, we explore two possible mechanisms. On the one hand, old authors with a high degree have a larger probability of attracting external new authors. On the other hand, old authors with a high degree have a larger probability of forming internal new collaborations. In Fig 5 and Fig 6, we calculate the probability of collaborating with external new authors and forming new internal links among old authors. The slope of the dashed line corresponds to the exponent of power law distribution. The slope values are positive, which signifies that authors with larger k are more likely to connect with new authors (Fig 5). From 2020 to 2021, the increase rates in the probability of attracting newcomers increases for top authors with more collaborators, indicated by the slope changes. In the first quarter of 2021, the second quarter of 2021 and the fourth quarter of 2021, the slopes are above 1. On the whole, the inequality of newcomer distribution is aggravated until 2021. Similarly, the slope values are positive in Fig 6, which indicates that authors with larger k are more likely to publish COVID-19 papers together. But the difference is the increase rate in the probability of building new connections among old nodes with high degree decreases from 2020 to 2021. In the second quarter of 2020, the third quarter of 2020 and the fourth quarter of 2020, the slopes are above 1.

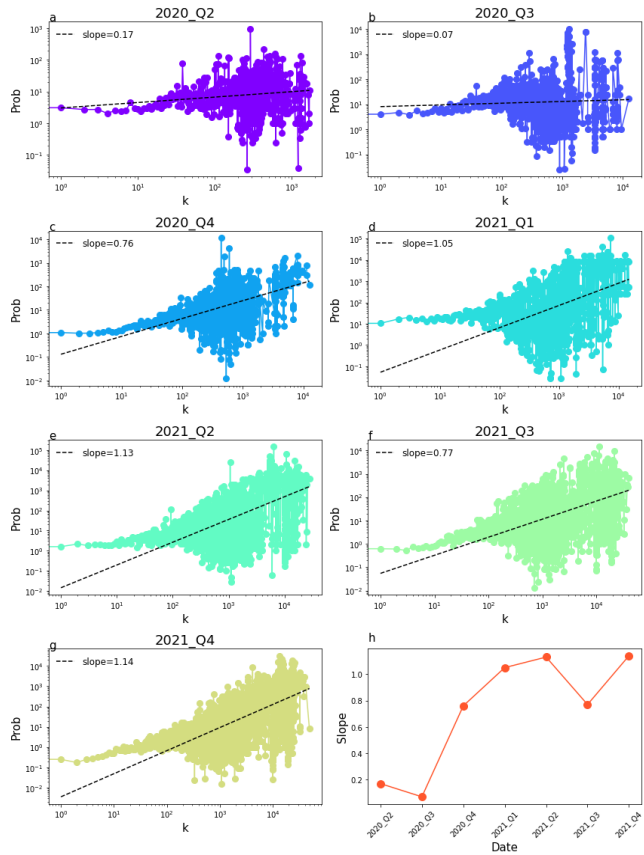

**Fig 5. The probability of attracting newcomers for existing COVID-19 authors before the given time. a-g.** The plot is drew using a logarithmic scale for both the x-axis and the y-axis. The x-axis and y-axis is calculated as Fig 2d. We fit the increasing trend with dashed lines and calculate the slope. **h.** The changes of slopes corresponding to a-g.

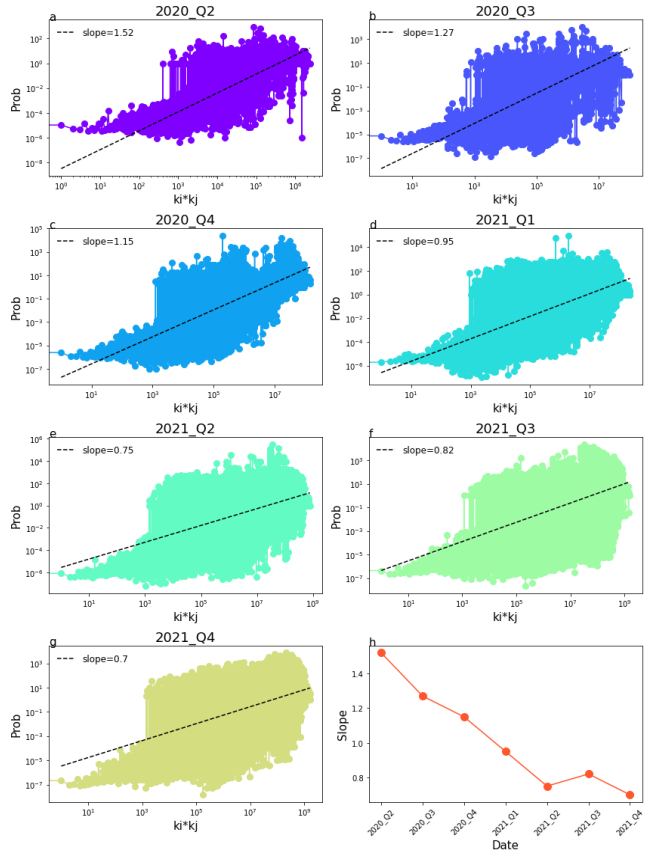

**Fig 6. The probability of forming new links among existing COVID-19 authors before the given time. a-g.** The plot is drawn using a logarithmic scale for both the x-axis and the y-axis. The x-axis and y-axis is calculated as Fig 2e. We fit the increasing trend with dashed lines and calculate the slope. **h.** The changes of slopes corresponding to a-g.

## 5 Conclusion

The current COVID-19 pandemic has caused a huge economic loss with the record high unemployment, the collapse of industry giants (e.g., large retails), the bankruptcy of small and medium-sized businesses, and spiral decline in spending, traveling, producing, and servicing. The whole world is on the pause button, and the world economy is on a spiral downturn. Uncertainty lies ahead of us. With our ever-growing highly connected world, simple infectious diseases can rapidly transform into pandemics, the threat and damage of future infectious diseases can be immense. Understanding scientific activities in the current and past pandemics/epidemics can help us identify patterns and pinpoint wrong-doings. This paper conducts preliminary research about the connectivity of scientific activities of COVID-19 authors who are working at the frontlines to fight against COVID-19. In addition to describing the static topology of the COVID-19 collaboration network, we also reveal the dynamic evolution of the network. We found that COVID pandemic alters the research collaboration trends that star scientists are starting to collaborate more with newcomers, but less with their peers, which, in certain way, reduces the collaboration disparity.